\documentclass[twocolumn]{aastex63}
\usepackage{amsmath}
\usepackage{comment}
\usepackage{graphicx}
\usepackage{xcolor}
\usepackage{xparse,xcoffins}

\newcommand{\be}{\begin{equation}}
\newcommand{\ee}{\end{equation}}
\newcommand{\Is}{\tilde{I}^2_s}
\newcommand{\rh}{\rho_{\rm{H}2}}
\usepackage{tgtermes}

\newcommand{\ci}{\ion{C}{1}}
\newcommand{\cii}{\ion{C}{2}}

\shorttitle{LIM Model Dependence}
\shortauthors{Breysse et al.}

\begin{document}

\title{On estimating the cosmic molecular gas density from CO Line Intensity Mapping observations}
\correspondingauthor{Patrick C. Breysse}
\email{pb2555@nyu.edu}

\author{Patrick C. Breysse}
\affiliation{Center for Cosmology and Particle Physics, Department of physics, New York University, 726 Broadway, New York, NY, 10003, U.S.A.}

\author{Shengqi Yang}
\affiliation{Center for Cosmology and Particle Physics, Department of physics, New York University, 726 Broadway, New York, NY, 10003, U.S.A.}

\author{Rachel S. Somerville}
\affiliation{Center for Computational Astrophysics, Flatiron institute, New York, NY 10010, U.S.A.}
\author{Anthony R. Pullen}
\affiliation{Center for Cosmology and Particle Physics, Department of physics, New York University, 726 Broadway, New York, NY, 10003, U.S.A.}
\affiliation{Center for Computational Astrophysics, Flatiron institute, New York, NY 10010, U.S.A.}
\author{Gerg\"o Popping}
\affiliation{European Southern Observatory, Karl-Schwarzschild-Strasse 2, D-85748, Garching, Germany}
\author{Abhishek S. Maniyar}
\affiliation{Center for Cosmology and Particle Physics, Department of physics, New York University, 726 Broadway, New York, NY, 10003, U.S.A.}

\begin{abstract}
The Millimeter-wave Intensity Mapping Experiment (mmIME) recently reported a detection of excess spatial fluctuations at a wavelength of 3 mm, which can be attributed to unresolved emission of several CO rotational transitions between $z\sim1-5$.  We study the implications of this data for the high-redshift interstellar medium using a suite of state-of-the-art semianalytic simulations which have successfully reproduced many other sub-millimeter line observations across the relevant redshift range.  We find that the semianalytic predictions are mildly in tension with the mmIME result, with a predicted CO power $\sim3.5\sigma$ below what was observed.  We explore some simple modifications to the models which could resolve this tension.  Increasing the molecular gas abundance at the relevant redshifts to $\sim10^8\ M_\odot\ \rm{Mpc}^{-3}$, a value well above that obtained from directly imaged sources, would resolve the discrepancy, as would assuming a CO-$H_2$ conversion factor $\alpha_{\rm{CO}}$ of $\sim1.5\ M_{\odot}$ K$^{-1}$ $(\rm{km}/\rm{s})^{-1}$ pc$^{2}$, a value somewhat lower than is commonly assumed.  We go on to demonstrate that these conclusions are quite sensitive to the detailed assumptions of our simulations, highlighting the need for more careful modeling efforts as more intensity mapping data become available.
\end{abstract}


\section{Introduction}
\label{sec:intro}

As the primary fuel for star formation, cold molecular gas plays a critical role in galaxy evolution.  From the first detections of molecular gas outside of our own Milky Way \citep{Rickard1975,Rickard1977,Solomon1975} there has been an explosion of observations targeting extragalactic molecular gas \citep[see reviews by][]{Carilli2013,Combes2018,Tacconi2020}.  Paralleling and informing similar observations of cosmic star formation \citep{Madau2014}, a consensus picture has emerged of the cosmic molecular gas history, with the overall abundance of molecular gas rising to a peak at redshift $z\sim2$ before declining to the present day \citep[see, e.g.][]{Walter2020}.

However, this wealth of observational data share one key limitation, that they can only access sources bright enough to detect individually, usually through their dust continuum or through rotational transitions of carbon monoxide.  Molecular gas surveys to date have generally applied one of two strategies, either directly targeting a sample of galaxies selected based on other observations \citep[e.g.][]{Tacconi2013,Freundlich2019}, or blindly scanning a field and reporting candidates above a certain detection threshold \citep[e.g.][]{Walter2014,Pavesi2018,Gonzlez2019}. Both of these methods will by their nature detect only a limited number of sources, and those sources will by definition be biased towards the brightest objects, which may or may not be representative of the full population.

Recently, a third observational technique known as line intensity mapping (LIM) has garnered attention as a complement to these approaches.  Intensity mapping does not aim to directly image individual objects, but instead seeks to make a statistical observation of the aggregate emission from many unresolved objects \citep[see][for a review]{Kovetz2017}.  By targeting spectrally narrow line emission, the target redshift can be finely selected by modifying the observing frequency, enabling tomographic measurements.  For molecular gas studies, LIM observations of CO lines are sensitive to fainter galaxies inaccessible to direct imaging, allowing population-wide examination of this crucial ISM component.  First discussed in \citet{Righi2008}, this application of LIM has garnered significant interest \citep{Lidz2011,Pullen2013,Breysse2014,Breysse2016,Breysse2017,Fonseca2017,Padmanabhan2018,Breysse2019,Chung2019,Dizgah2019}, with a number of experimental efforts planned or in progress \citep{Li2016,Lagache2018,Stacey2018,Sun2020,Cataldo2021}.  Two of these experiments, the CO Power Spectrum Survey \citep[COPSS,][]{Keating2016} and the Millimeter Intensity Mapping Experiment \citep[mmIME,][hereafter K20]{Keating2020}, have reported 2 and $4-\sigma$ detections of unresolved CO emission respectively.

Interpreting the results of CO observations, whether direct or LIM, is a challenging modeling task.  The conversion factor $\alpha_{\rm{CO}}$ between CO luminosity and molecular gas mass is known to depend on metallicity, density, and the level of excitation of the CO-emitting gas \citep{Bolatto2013,Tacconi2020}.  Thus the molecular abundance inferred from a measurement depends sensitively on modeling assumptions.  LIM measurements add an extra level of complexity to the modeling task.  The very thing which gives LIM surveys their sensitivity to faint populations, that they do not seek to detect individual sources, means that LIM observations are only able to measure quantities integrated over the entire galaxy population.  Thus, in addition to modeling the CO-gas conversion for individual sources, a LIM measurement also needs to model how that conversion is distributed across all galaxies, including those too faint to detect individually.

In this work, we study how these modeling choices affect the interpretation of modern CO intensity maps.  We will focus on the aforementioned mmIME survey as a test case.  K20 reported a detection of aggregate CO emission at a wavelength of 3 mm using data from the Atacama Large Millimeter Array (ALMA).  This observed wavelength contains emission from several CO transitions at different redshifts, and thus the claimed detection contains the sum of contributions from each of these lines.  K20 then made use of several scaling relations from the literature to convert their CO detection into a cosmic molecular gas history, finding an abundance on the high end of but consistent with results from direct imaging surveys.  As we will show, these results are quite sensitive to the details of the assumed scaling relations.  The relations used in K20 are all themselves calibrated to relatively small populations of directly-detected galaxies, coming from a variety of observed bands and redshifts.  They are all empirical in nature, and thus there is no underlying physics behind the various steps in the conversion.  Here we will seek to place this interpretation on a more physically grounded footing by employing a semi-analytic model (SAM) of galaxy evolution combined with a line spectral synthesis code.

The primary observational quantity reported by the mmIME survey is the total spectral shot power $\Is$, which can be directly related to integrals over the CO luminosity function.  The observed emission comes from several overlapping CO transitions at different redshifts, so one must model the redshift distribution of the CO transitions to assign a fraction of the total power to each line.  Once this is done, each line must be converted to the equivalent CO(1-0) luminosity, and finally to a molecular gas abundance through an assumed $\alpha_{\rm{CO}}$value.  Each step must include an assumption about the mass dependence of the relevant conversion.  As the mmIME result provides only a single number to constrain this wide model space, the approach taken in K20 is to effectively assume that the shape of all of these distributions is known, and to fit an overall amplitude to match the observed shot power.  We will follow a similar procedure, but we will obtain all of the relevant distributions and scalings from the suite of SAMs described in \citet[][hereafter Y20]{Yang2020}.  The SAM procedure seeks to predict CO luminosity in a self-consistent, physically-motivated fashion, and is calibrated to a wide variety of empirical measurements of multiple lines.  This should give them broader applicability and more robustness than any single empirical scaling.  We use a two-step process consisting of a galaxy evolution SAM developed by the Santa Cruz group \citep{2015MNRAS.453.4337S} with an additional subgrid model which predicts sub-mm line emission based on galaxy properties \citep{2019MNRAS.482.4906P}.  We will refer to this suite of models as the "SAM+sub-mm SAM". 

Our primary result is a modest tension between the predictions of the semi-analytic models and the overall power detected by mmIME.  While the scalings used in K20 for $L_{CO}$ predict a value of $\Is$ quite close to the mmIME measurement, the raw SAM+sub-mm SAM prediction is lower by an order of magnitude, corresponding to a $\sim3.5-\sigma$ discrepancy.  We find that, given this tension, the same amplitude-scaling procedure applied in K20 but with a SAM+sub-mm-SAM-informed scaling yields a correspondingly high cosmic molecular gas abundance.  Taken at face value, this would suggest that the LIM survey sees dramatically more molecular gas in the universe than has been mapped directly\footnote{Most direct CO surveys estimate the molecular gas abundance using only directly detected sources, without extrapolating the luminosity function to fainter emitters \citep[see, e.g.][]{Walter2014,Decarli2019,Lenkic2020}}.  This result is not outside the realm of possibility, as LIM surveys by their nature are sensitive to more emission than traditional surveys, but one must be careful in drawing such a strong conclusion from limited data.  To demonstrate this, we go on to examine how varying the model assumptions affects the molecular gas estimates.  We find that the tension can be reduced by assuming a conventional molecular gas history and reducing the average $\alpha_{\rm{CO}}$ value below what is generally assumed for high-redshift galaxies.  Similarly, while it does not remove the tension in overall amplitude, changing the assumed density profile of gas clouds in the SAM+sub-mm SAM lowers the amount of molecular gas needed to resolve the tension back down near the level obtained by the original mmIME analysis.  Finally, we substitute back in various pieces of the empirical scalings used in K20 and find that the scatter in the resulting molecular gas abundance is significantly larger than the statistical error bars on the measurement, pointing to the need for more data and more careful modeling to resolve this tension.

The organization of this paper is as follows:  Section \ref{sec:SAM_intro} reviews the details of the SAM+sub-mm SAM models used in this analysis.  Section \ref{sec:method} describes how both the K20 scalings and the SAM+sub-mm SAM are used to predict the mmIME signal and illustrates the basic tension described above. Section \ref{sec:gas_connection} then reviews the K20 procedure for converting the mmIME measurement to a molecular gas abundance, and describes the modified version of that method used here, with the comparison of the results of these methods described in Section \ref{sec:results}.  Section \ref{sec:variations} describes the effects of varying our model assumptions.  We discuss our results and how they will improve in the future in Section \ref{sec:discussion} and conclude in Section \ref{sec:conclusion}.  Throughout this paper we assume a flat $\Lambda$CDM cosmology consistent with the Planck 2018 results \citep{Planck2020}.

\section{Semi-analytic models and sub-mm SAM}\label{sec:SAM_intro}
The sample of DM halos and galaxies adopted in this paper are provided by the Y20 cosmological mock lightcone. Here we briefly summarize the lightcone simulation framework and parameters. We refer readers to Y20 for a more detailed description.\par 
The Y20 workflow can be summarized as a three-step process. In the first step, dark matter halos are selected from an N-body simulation along a past lightcone, as described in  \citet{Somerville:2021}, using the \texttt{lightcone} package provided by P. Behroozi \citep{Behroozi:2019}. 
We choose the Small MultiDark-Planck (SMDPL) N-Body simulation \citep{2016MNRAS.457.4340K} to provide the DM halo information for a 2 deg$^2$ wide and $0\leq z\leq10$ long cosmological lightcone. Since the simulation cube volume of SMDPL is $(400 \mathrm{Mpc/h})^3$, greater than the volume of the target lightcone $5.7\times10^7 (\mathrm{Mpc/h})^3$, the statistical independence among different regions within the lightcone is in principle guaranteed. Moreover, the $10^{10} M_\odot$ halo mass resolution of SMDPL is fine enough to ensure that massive halos that can retain significant gas reservoirs and make significant contribution to the line intensity mapping (LIM) statistics are not omitted.\par 
In the second step, a SAM developed by the Santa Cruz group \citep{1999MNRAS.310.1087S,2008MNRAS.391..481S,2012MNRAS.423.1992S,2014MNRAS.444..942P,2014MNRAS.442.2398P,2015MNRAS.453.4337S,Somerville:2021} is used to simulate the DM halo merger history as well as the formation and evolution of galaxies. The Santa Cruz SAM simulates the halo merger history using a method based on the extended Press-Schechter (EPS) formalism \citep{1999MNRAS.305....1S} or extracts the DM halo merger trees directly from N-Body simulations. We adopt the EPS method in this work as it allows us to resolve halos down to much lower masses. It also replaces the computationally expensive hydrodynamic simulations with simplified but physically motivated treatments, covering the galaxy merging, gas heating and cooling, ISM gas partitioning, photoionization squelching, and various feedback processes. 

The Santa Cruz SAM we use in this work has successfully reproduced various UV/optical galaxy observations up to $z=8$ \citep{2012MNRAS.423.1992S,2015MNRAS.453.4337S,2019MNRAS.483.2983Y,2019MNRAS.490.2855Y,Somerville:2021}.  However, it is important to keep in mind that, like all current galaxy formation models, there are still many uncertainties in the key physical processes implemented within the Santa Cruz SAMs, and some of their predictions are in tension with some observations \citep[see e.g.][]{2019ApJ...882..137P,Somerville:2021}.
\par

In the last step, a sub-resolution recipe developed by \citep[][hereafter P19]{2019MNRAS.482.4906P} is used to simultaneously model the [\cii], CO, and [\ci] emission of each of the simulated galaxies, and we refer this model as the sub-mm SAM in the following. The Santa Cruz SAM divides each simulated galaxy into multiple annuli, and uses the predicted gas surface density, metallicity, and SFR in each annulus to predict the fraction of the ISM that is in the form of molecular, atomic, and ionized gas. The sub-mm SAM then randomly selects molecular cloud (MC) masses in each annulus from a power-law MC mass function until the total mass of MCs reaches the molecular gas mass predicted by the SAM. The sub-mm SAM then assumes a MC radial density profile and further grids each MC into 25 zones. In each zone it estimates the emission luminosity of multiple sub-mm lines with a linearly interpolated lookup table given by the line spectral synthesis code \textsc{DESPOTIC} \citep{2014MNRAS.437.1662K}. Finally, the galaxy-wide [\cii], CO, and [\ci] luminosities are computed by summing over the emission contributed by all the MC zones in all annuli. 

The sub-mm SAM has the advantage of simultaneously and self-consistently estimating multiple sub-millimeter emission lines and their correlations, based on a set of properties predicted by a cosmological galaxy formation model. However, the sub-mm SAM introduces free parameters to describe the MC mass distribution, radial density profile, and the external radiation fields. Currently, calibration is performed roughly through manual parameter tuning. Particularly relevant to this work, P19 found that the choice of MC radial density profile has a significant influence on the resulting sub-mm line emission statistics. GP19 showed that the sub-mm SAM predictions with an assumed Plummer density profile agree best with the observed line luminosity versus galaxy SFR relations that were available at publication (2018). However, Y20 found that the original sub-mm SAM prediction is about a factor of three lower than more recent ALMA [\cii] measurements \citep{2018MNRAS.481.1976Z,2020A&A...643A...2B}. \cite{2020ApJ...890...24V} also suggests that the sub-mm SAM [\ci] luminosity versus IR luminosity predictions at high redshifts are significantly lower than the observations. Y20 showed that using a power-law MC density profile can largely alleviate those tensions. In this work we therefore adopt the power-law MC radial density profile as our baseline assumption, and we discuss how much the MC density profile assumption influences molecular gas density constraints in \S \ref{sec:Plummer}.

\section{Estimating LIM shot power from line emission models}
\label{sec:method}

In this section, we outline the formalism used to connect galactic line emission to the spectral shot power in an intensity map.  We will describe how the two modeling routines we consider here predict the amplitude of this shot power and compare their predictions to the value observed by mmIME to illustrate the tension between the semi-analytic models and the LIM data.

K20 report their baseline result in the form of a single measurement of the total spectral shot power $\Is$ in a band centered at a wavelength of $\sim3$ mm.  On the small ($k\sim10\ h$ Mpc$^{-1}$ or more) spatial scales relevant to this measurement, the total power is a weighted sum of the individual shot noise levels $P_{\rm{shot}}$ for every emission line that falls into this frequency band, given by
\be
\Is(\nu_{\rm{obs}})=\sum_{\rm{all lines}}\frac{P_{\rm{shot,line}}(z)}{X^2(z)Y(z,\nu_{r,\rm{line}})}.
\label{Is}
\ee
Each line is observed at a redshift $z=\nu_r/\nu_{\rm{obs}}$ assuming a rest frequency $\nu_r$ and an observed frequency $\nu_{\rm{obs}}$.  The power in each line is rescaled to project the spatial dimensions at its emission redshift to a common observed frame, in the angular direction by $X(z)=D_M(z)$ where $D_M$ is the comoving radial distance to redshift $z$, and in the line-of-sight direction by $Y(z,\nu_r)=c(1+z)^2/H_0E(z)\nu_r$, where $c$ is the speed of light, $H_0$ is the Hubble constant at $z=0$, and $E(z)$ is the dimensionless Hubble parameter.  In the band observed in K20, the observed emission is dominated by the CO rotational series, with the dominant emission coming from the CO(2-1)-(5-4) transitions from redshifts $\sim1-5$.

The shot noise $P_{\rm{shot}}$ that appears in Eq. (\ref{Is}) is a quantity which arises in LIM measurements due to the discrete nature of most line-emitting sources.  CO emission arises virtually entirely from dense gas within the ISM of star-forming galaxies, which on cosmological scales effectively form a population of point-source emitters.  On large scales, the clustering of dark matter halos causes the locations of these emitters to be correlated, but on small scales their power spectra are dominated by random, approximately-Poisson density fluctuations.  This gives rise to a scale-independent power spectrum on small scales with an amplitude
\begin{multline}
P_{\rm{shot}}(z)=C_{LI}^2(z)e^{[\sigma\ln(10)]^2} \\ \times\int_{M_{\rm{min}}}^{\infty} L_0^2(M)f_{\rm{duty}}(M)\frac{dn}{dM}dM,
\label{Pshot}
\end{multline}
\citep{Lidz2011}.  This differs from the shot noise in a spectroscopic galaxy survey as each galaxy is weighted by its line luminosity.  In Eq. (\ref{Pshot}), $L_0(M)$ is the mean line luminosity of halos with mass $M$, which is then averaged over the halo mass function $dn/dM$.  We assume the \citet{Tinker2008} halo mass function throughout this work.  The quantity
\be
C_{LI}(z)=\frac{c}{4\pi\nu_r H(z)}
\ee
converts from luminosity density to intensity units.  As is common in the literature, we assume that halos with masses below $M_{\rm{min}}=10^{10}\ M_{\rm{sun}}$ do not emit any appreciable CO, though our results are relatively insensitive to the exact value of this choice.

Many simpler LIM models assume that all halos of mass $M$ emit at exactly $L_0(M)$.  Here we relax this assumption in two important ways.  First, we fold the dependence of luminosity on other halo properties into a lognormal scatter around $L_0(M)$ of width $\sigma(M)$, as was first done in \citet{Li2016}.  Second, we assume that only a fraction $f_{\rm{duty}}(M)$ of all halos of a given mass are emitting at any given time \citep{Lidz2011,Pullen2013}.  This is particularly important at lower redshifts, where many galaxies may be quenched and emit effectively zero CO.  We have slightly generalized the $f_{\rm{duty}}$ factor here, both by allowing it to explicitly depend on halo mass, and by including both it and a scatter at the same time.  \citet{Keating2016} discussed accounting for both scatter and quenching with a single choice of $\sigma$, but we find that this does not provide an accurate description of the galaxy population produced by SAM.

We predict $P_{\rm{shot}}$ for the SAM+sub-mm SAM by binning all of the halos in our simulated catalog by halo mass.  We then account for quenching by removing all halos with a specific star formation rate (sSFR) below $1/3t_H(z)$, where sSFR is the ratio between star formation rate and halo mass and $t_H(z)$ is the Hubble time at redshift $z$.  The SAM+sub-mm SAM equivalent of $f_{\rm{duty}}$ is then the fraction of galaxies which remain in a mass bin after this cut.  Each mass bin then inherently includes a scatter about the mean halo mass, which arises from the variations in halo merger history within the ensemble of halos.  We can use this to estimate a lognormal scatter width $\sigma(M)$ for comparison with other models, but when we predict the shot power we can obtain a more accurate result by taking
\be
P^{\rm{SAM}}_{\rm{shot}}=C_{LI}^2(z)\int \left<L_0^2(M)\right>f_{\rm{duty}}(M)\frac{dn}{dM}dM,
\label{Pshot_SAM}
\ee
where $\left<L_0^2(M)\right>$ is obtained by interpolating the mean square CO luminosity in each mass bin (i.e., using the full predicted distribution of luminosities within each mass bin, instead of assuming that it has a log-normal form).  We could further improve the accuracy by removing the explicit $f_{\rm{duty}}$ and $dn/dM$ dependence from Eq. (\ref{Pshot_SAM}) and simply integrating over an interpolated luminosity function, but leaving the expression in this form lets us use the same mass function for all of our calculations, bringing us closer to an apples-to-apples comparison between the two approaches.

\begin{figure}
\centering
\includegraphics[width=\columnwidth]{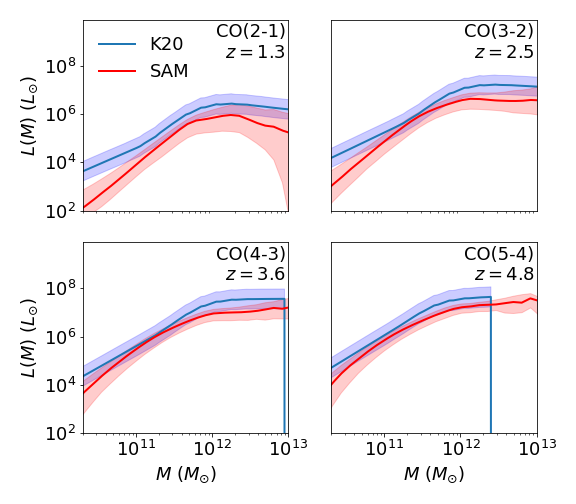}
\caption{Mass-luminosity relations for the four CO transitions we consider here assuming the modified Li model (blue) and the SAM prediction (red).  Shaded bands show the width of the $1-\sigma$ scatter.  The K20 models cut off at the maximum mass available in the \citet{Behroozi2013} star formation rate catalogs.}
\label{fig:LofM}
\end{figure}

\begin{figure}
\centering
\includegraphics[width=\columnwidth]{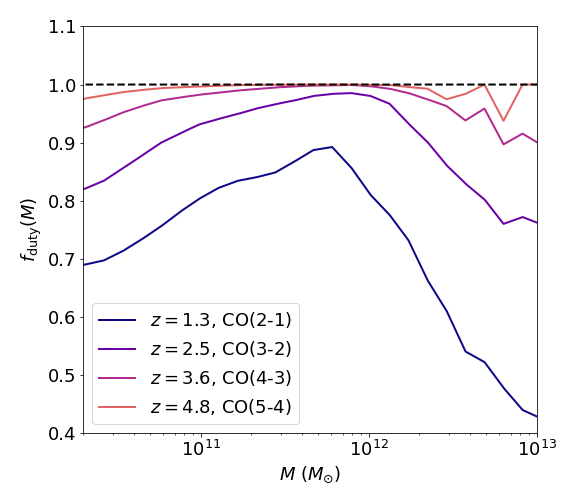}
\caption{Fraction $f_{\rm{duty}}$ of halos with non-negligible star formation as a function of halo mass at the redshifts of each of the four CO transitions we consider here inferred from the SAM+sub-mm SAM.  Though $f_{\rm{duty}}$ is a property of halos rather than specific CO transitions, we have labeled the relevant transition in the legend to track which $f_{\rm{duty}}$ curve is used for each line.}
\label{fig:fduty}
\end{figure}

Figure \ref{fig:LofM} shows the mean mass-luminosity relationships for the dominant CO transitions considered here.  As we are concerned with an observation at a single frequency, in this figure and for the rest of this text we will assume that each line is sourced only at the redshift corresponding to $\nu_{\rm{obs}}$, with CO(2-1) at an average redshift of 1.3, CO(3-2) at 2.5, CO(4-3) at 3.6, and CO(5-4) at 4.8.  Shaded regions around each curve in Fig. \ref{fig:LofM} denote the 68\% confidence interval of the scatter around the mean relation.  The higher redshift predictions for the K20 model cut off above some halo mass due to the mass limits of the \citet{Behroozi2013} star formation rate catalog.  Figure \ref{fig:fduty} shows the $f_{\rm{duty}}(M)$ values inferred from the SAM at the central redshifts where mmIME observes the four CO lines.  This value is quite close to unity at high redshift, but drops for lower-redshift lines as quenching effects become more important.

\begin{figure}
\centering
\includegraphics[width=\columnwidth]{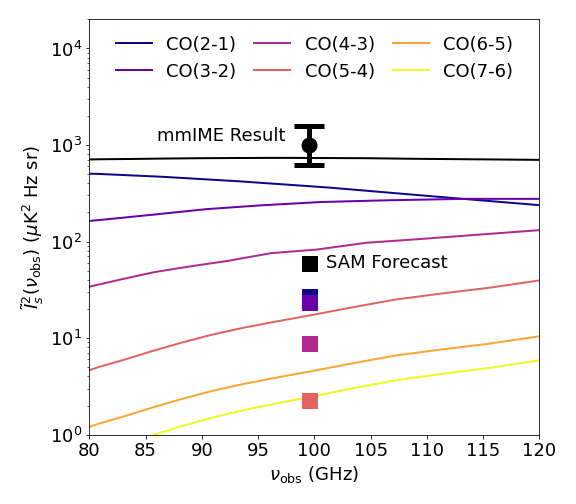}
\caption{Spectral shot power $I_s^2(\nu)$ for CO rotational transitions (colored curves/points) and the total power observed at a given frequency (black curves/points).  The circular point and error bars show the best fit value and 60\% confidence interval from K20.  Lines show the predictions as a function of observed frequency for the modified \citet{Li2016} model used in K20 (compare to their Fig. 7).  Squares show the values predicted by the SAM+sub-mm SAM.  Note that the semi-analytic prediction falls significantly below the mmIME measurement.}
\label{fig:Is}
\end{figure}

With these relations in hand, we can predict the total spectral shot power present in these two models and compare to the value reported in K20, as shown in Figure \ref{fig:Is}. As noted in K20, the modified \citet{Li2016} scaling model predicts a total shot power comparable to the mmIME measurement.  The physically-motivated semi-analytic models, on the other hand, produce a prediction a factor of $\sim20$ smaller, corresponding to a $\sim3.5\sigma$ discrepancy under the quoted mmIME sensitivity\footnote{The error quoted in K20 would correspond to a $\sim2.5\sigma$ discrepancy under the assumption of Gaussian uncertainty.  The full non-Gaussian likelihood leads to a stronger tension (G. K. Keating, pvt. communication)}.  This difference can be largely attributed to the nearly-universally lower amplitude of the $L_0(M)$ relationships seen in Fig. \ref{fig:LofM}.  The lowest-redshift CO lines make up a bit of power in the SAM+sub-mm SAM due to extra scatter about the mean relation, but this increase is offset by the inclusion of the sub-unity $f_{\rm{duty}}$ from Fig. \ref{fig:fduty}.  The separation between the semi-analytic prediction and the LIM observation seen in Fig. \ref{fig:Is} is the tension we seek to explore in the rest of this work.

\section{Connection to Molecular Gas}
\label{sec:gas_connection}

If we are to understand the consequences of this possible tension, we need to connect the integrated CO observations discussed above to the physical properties of the emitting galaxies.  For maps of CO emission, the most obvious quantity to explore is the molecular gas abundance, as we expect molecular clouds to source virtually all of the CO emission we detect.  K20 laid out a prescription for constraining the cosmic molecular gas history from the mmIME $I_s^2(\nu)$ measurement.  In this section, we will briefly review this procedure, then describe a more streamlined method enabled by the internal consistency of the SAM+sub-mm SAM catalog.

\subsection{Redshift Evolution}

The first hurdle we need to clear is to separate the single-frequency mmIME measurement into its contributions from the individual CO lines.  Many methods have been proposed in the literature for separating the signal from a target line from interloper emission \citep{Gong2014,Breysse2015,Lidz2016,Cheng2016,Sun2018,Cheng2020}.  Most of these tools, however, require either an additional cosmological tracer for masking or cross-correlation, or a larger, deeper map than those produced by mmIME.  Thus, K20 adopted a simpler estimate in which each line is assigned a fraction 
\be
f_{\rm{tot}}=\frac{\tilde{I}_{s,\rm{line}}^2}{\tilde{I}_s^2}
\ee
of the total observed shot power.  They then assigned the $f_{\rm{tot}}$ which would be obtained under the modified \citet{Li2016} model.  In other words, they assumed that the scaling model had the correct redshift evolution, and only the overall amplitude needed to be shifted to match the measured shot power.  

We can carry out the same prescription using our semi-analytically-derived relations, and calculate a predicted $f_{\rm{tot}}$ from the mass-luminosity relations plotted in Fig. \ref{fig:LofM}.  The results of this calculation are given in Table \ref{tab:ftot}.  One can immediately see that the two models assign similar fractional contributions to each line.  This may be due to both models having been calibrated on similar multi-line source data.
Both models assign $<1\%$ of the total signal to lines with higher frequency than CO(5-4), so we will adopt the same assumption as K20 and neglect any contribution from higher-order CO transitions.

\begin{table}
	\centering
	\caption{Predicted contributions to the total shot power $\Is$ from each line considered here, under the modified \citet{Li2016} model used and the SAM+sub-mm SAM model.  All $\Is$ values have dimension $\mu$K$^2$ Hz sr.}
	\label{tab:ftot}
	\begin{tabular}{c|c|cc|cc} 
		\hline
		Transition & $\left<z\right>$ & K20 $\Is$ & SAM $\Is$ & K20 $f_{\rm{tot}}$ & SAM $f_{\rm{tot}}$\\
		\hline
		CO(2-1) & 1.3 & 340 & 27.5 & 0.43 & 0.44 \\
		CO(3-2) & 2.5 & 310 & 23.5 & 0.4 & 0.38 \\
		CO(4-3) & 3.6 & 120 & 8.8 & 0.14 & 0.14 \\
		CO(5-4) & 4.8 & 30 & 2.3 & 0.03 & 0.04 \\
		\hline
		Total & -- & 790 & 62.1 & 1.00 & 1.00\\
		\hline
	\end{tabular}
\end{table}

This method of separating interloper lines is obviously highly simplified, and despite the similarity between these two models the data would certainly permit a substantially different redshift evolution.  Furthermore, we see in Table 4 that a $\sim1\%$ change in the total intensity corresponds to a $\sim30\%$ increase in the CO(5-4) shot power.  This is a problem intrinsic to any measurement of a faint signal overlapping with brighter emission.  We clearly should not trust these models to be accurate to the percent level, so especially the highest-redshift estimates made under these assumptions should be taken with a grain of salt.  However, our purpose here is a rough exploration of this pair of models, not a precision measurement.  Between this and the fact that is difficult to derive a better redshift distribution without further measurements, we leave a more detailed exploration of the redshift evolution of the LIM signal to future work.

\subsection{Scaling Models}
\label{sec:scalings}

Now that we have an estimate for at least one moment of the CO luminosity function in our four redshift bins, we can convert that value into a measurement of the cosmic molecular gas abundance $\rh$.  First we will review the method used in K20 for this conversion.  Typically, the molecular gas content $M_{\rm{H2}}$ of a galaxy is related to its CO luminosity by the parameter
\be
\alpha_{\rm{CO}}=\frac{M_{\rm{H2}}}{L'_{\rm{CO(1-0)}}},
\label{aCO}
\ee
where the observer-unit CO luminosity $L'$ is related to the physical version by
\be
\frac{L_{\rm{CO(J)}}}{L_{\odot}}=4.9\times10^{-5}\left(\frac{\nu_J}{\nu_{\rm{(1-0)}}}\right)^3\left(\frac{L'_{\rm{CO(J)}}}{\rm{K\ km\ s^{-1}\ pc^2}}\right).
\label{Lprime}
\ee
K20 assume a typical extragalactic value of $\alpha_{\rm{CO}}=3.6\ M_{\odot}$ (K km s$^{-1}$ pc$^{2}$)$^{-1}$, with wider estimates ranging from a Milky Way-like $\alpha_{\rm{CO}}=4.3\ M_{\odot}$ (K km s$^{-1}$ pc$^{2}$)$^{-1}$ down to a ULIRG-like $\alpha_{\rm{CO}}=0.8\ M_{\odot}$ (K km s$^{-1}$ pc$^{2}$)$^{-1}$.

Since $\alpha_{\rm{CO}}$ is typically reported relative to CO(1-0) luminosity, it is necessary to convert the higher-J CO intensities predicted here to their CO(1-0) equivalents.  K20 adopt the mean line ratios $r_{J,1}=L'_J/L'_{(1-0)}$ from a sample of $z=1.5$ optically-selected galaxies \citep{Daddi2015} and average them over the galaxy population by assuming
\be
P_{\rm{shot,J}}=r_{J,1}^2P_{\rm{shot,CO(1-0)}},
\label{rJ}
\ee
where $r_{J,1}=0.76\pm0.09$, $0.42\pm0.07$, $0.31\pm0.06$, and $0.23\pm0.04$ for the CO(2-1), (3-2), (4-3), and (5-4) transitions respectively.  

Finally, we need a prescription for fitting a model for $L_{\rm{CO,J}}$ to the measured $\Is$. As we have only a single number to base this on, we can handle only a single independent parameter in our fitting.  K20 assume a broken power law mass-luminosity relation for this step parameterized by
\be
L_{\rm{CO}}(M)=\begin{cases} 
      A_{\rm{CO}}\frac{M^2}{M_0} & M\leq M_0 \\
      A_{\rm{CO}}M_0 & M\geq M_0 
   \end{cases},
   \label{newscaling}
\ee
where $M_0=10^{12}\ M_{\odot}/h$ is the location of the turnover and the overall amplitude $A_{\rm{CO}}$ is the free parameter used for the fit.  When applying Eq. (\ref{newscaling}), the K20 model assumes the default mass-independent scatter of $\sigma=0.37$ dex from \citet{Li2016}  As with all of the K20 scaling relations, we continue to assume $f_{\rm{duty}}=1$.

Thus, the overall procedure to connect the measured $\Is$ to $\rho_{\rm{H2}}$ goes as follows:
\begin{itemize}
\item Find the value of $A_{\rm{CO}}$ in Eq. (\ref{newscaling}) which best fits the measured shot power $\Is$.
\item Convert the resulting CO luminosity to the CO(1-0) equivalent using Eq. (\ref{rJ}).
\item Convert the CO(1-0) luminosity to $M_{\rm{H2}}$ assuming a constant $\alpha_{\rm{CO}}$
\item Integrate the resulting $M_{\rm{H2}}(M)$ over a halo mass function to get a total estimated $\rh$.
\end{itemize}

\subsection{Alternate procedure using semi-analytic models}
\label{sec:sam-scale}
The full range of detailed galaxy properties available in the SAM+sub-mm SAM enable a more nuanced procedure for constraining $\rh$ from $\Is$.  The above scaling approach necessarily assumes that most of its relevant quantities are both mass and redshift independent.  Due to the sparsity of empirical data, many of the scalings are calibrated at different redshifts than the mmIME observation.  Furthermore, all of the scalings should have some galaxy-to-galaxy scatter about the given mean relation, just like that of the the mass-luminosity relation discussed previously.  Scatter in one relation can easily be correlated with scatter in another, further complicating attempts to model these relationships analytically.  

With our simulated halos in hand, we can calculate the full mass- and redshift dependent forms of Eqs. (\ref{aCO}-\ref{newscaling}) based on the SAM.  However, while we will make use of these later, we do not actually need them to get a $\rho_{\rm{H2}}$ measurement.  Following the broad strokes of the K20 model, we will assume we know the shape of the $L_0(M)$ relation from the SAM results and fit for an overall amplitude offset $A^{\rm{SAM}}_{\rm{CO}}$ such that
\be
L_{0,\rm{fit}}(M)=A^{\rm{SAM}}_{\rm{CO}}L_0(M).
\ee
In other words, we want to find the value of $A^{\rm{SAM}}_{\rm{CO}}$ that brings the SAM+sub-mm SAM forecast for $\Is$ into agreement with the mmIME result plotted in Fig. \ref{fig:Is}, assuming the semianalytic predictions for the mean and scatter on $L_0(M)$ as well as $f_{\rm{duty}}$.

Since we know the molecular content of each SAM galaxy, we know the value of the total cosmic molecular gas density $\rho_{\rm{H2},0}$ that corresponds to the default SAM+sub-mm SAM model results.  By definition then, $A^{\rm{SAM}}_{\rm{CO}}=1$ corresponds to a universe with $\rh=\rho_{\rm{H2},0}$, and the total shot power $\Is$ scales as $(A^{\rm{SAM}}_{\rm{CO}})^2$. Since we are rescaling the CO luminosities linearly, we can get a K20-equivalent $\rh$ measurement simply from
\be
\rho_{\rm{H2}}=A^{\rm{SAM}}_{\rm{CO}}\rho_{\rm{H2},0}.
\label{rhoSAM}
\ee
Note that, in the SAM, we of course know the predicted value of both the conversion from any excitation state CO line to the CO(1-0) line luminosity, and we know the value of $\alpha_{\rm{CO}}$ \emph{for each galaxy}. Indeed, these conversions vary galaxy by galaxy and also depend on redshift indirectly, due to changes in the ISM properties and the CMB radiation field strength \citep[see e.g.][]{Popping:2016}. With the procedure described above, we are assuming a population-averaged effective conversion between observed line emission and CO(1-0), and similarly a population averaged value of $\alpha_{\rm{CO}}$. This implicitly assumes that these conversion factors are correctly predicted by the SAM, \emph{and} that the SAM correctly predicts the galaxy population properties that are relevant for determining them. 

\section{Results}
\label{sec:results}

\begin{figure}
\centering
\includegraphics[width=\columnwidth]{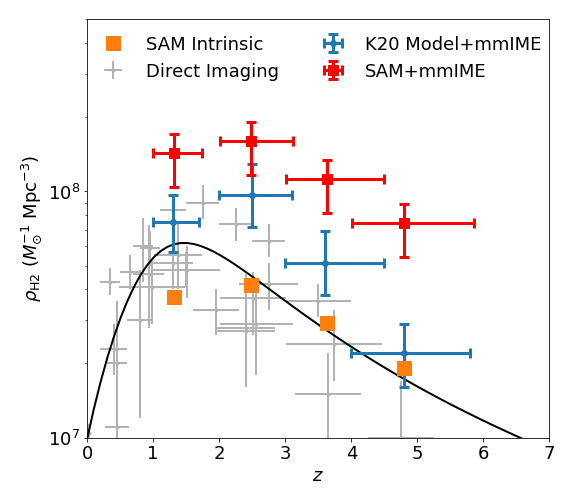}
\caption{Measurements of the cosmic molecular gas history.  Grey points show direct measurements collected in W20; the black line shows a fit to these points.  Blue points are the values quoted in K20 assuming  empirical scalings.  Orange squares show the intrinsic values of $\rho_{\rm{H2}}$ in the SAM, and red points show the values inferred from the mmIME measurement using conversions based on the SAM. Note that, because all of the LIM measurements are derived from a single measured data point, all of the errors on the red and blue points are perfectly correlated.}
\label{fig:rhoH2}
\end{figure}

Here we present the molecular gas history inferred by using the procedure based on the semi-analytic mocks, as outlined in \S\ref{sec:sam-scale}, in place of the K20 scaling relations as described in \S\ref{sec:scalings}.  Figure \ref{fig:rhoH2} shows the resulting $\rh(z)$ estimate compared to that obtained in the original mmIME analysis.  We further compare those measurements to a compilation of direct-imaging results from \citet[][hereafter W20]{Walter2020}, which includes CO-based measurements from the xCOLD GASS \citep{Fletcher2021}, ASPECS \citep{Decarli2020}, COLDz \citep{Riechers2019} and VLASPECS surveys \citep{Riechers2020}, along with several dust continuum observations \citep{Berta2013,Scoville2017,Magnelli2020}.  The black line shows a fit to these direct observations.  The original mmIME results quoted in K20 are broadly consistent with the direct imaged molecular gas surveys, if slightly on the high side.  The values of $\rho_{\rm{H2}}$ obtained by directly integrating the intrinsic molecular gas masses in all galaxies in the SAM are also consistent with the values estimated from direct imaging surveys. 

Due to the discrepancy between the predicted and measured $\Is$, however, the SAM+sub-mm SAM requires an $A^{\rm{SAM}}_{\rm{CO}}$ of $\sim4$ to match the mmIME $\Is$ measurement.  Thus, we find that using the scalings derived from the SAM and the mmIME measurement of $\Is$, instead of the K20 scalings, increases the inferred value of $\rh$ from a value consistent with the direct measurements to one substantially above them. 

It should be noted that, again because we are only fitting a single data point, the four redshift bins plotted in Fig. \ref{fig:rhoH2} are all exactly correlated.  Thus what we are seeing is the same $\sim3.5\sigma$ tension from Fig. \ref{fig:Is} being carried over into the estimates of  $\rh$.  What we have effectively done is assumed that the relationship between H$_2$ mass and CO luminosity is exactly known from the SAM+sub-mm SAM and linearly increased the amount of molecular gas to needed to match the mmIME $\Is$ value.  

If all of the above assumptions are taken at face value, this would seem to indicate that mmIME has seen a substantial reservoir of high-redshift molecular gas that has not been seen by traditional surveys.  This is not implausible on the surface, as one of the primary motivators for intensity mapping surveys like mmIME is to map the properties of sources below the detection thresholds of individual images.  

It is clearly premature to make such a strong statement though, at least without substantial caveats.  A single data point in modest tension with a single model is not enough justification to claim a detection of a factor of several increase in the molecular gas content of the universe, particularly when modeling a system as complicated and poorly understood as the high-redshift interstellar medium.  We will devote the next section to an exploration of some of the ways in which the results of this analysis change depending on the model assumptions we adopt.

\section{Varying the Model}
\label{sec:variations}

As noted in \S\ref{sec:SAM_intro}, both the SAM and sub-mm SAM contain free parameters that characterize the physical processes in the galaxy formation model (e.g. star formation efficiency, stellar feedback efficiency, AGN feedback efficiency, etc) and the sub-grid properties of the molecular clouds that populate the ISM and give rise to the observed CO emission (cloud mass function, radial profile, etc). These parameters have, up to now, been manually tuned to match the properties of individual observed sources like the ones used to derive the W20 direct estimate of $\rh$. In principle, we could try varying these parameters, perhaps using something like a Monte Carlo Markov Chain sampler, to explore whether there are parts of the complex and high-dimensional parameter space that can satisfy the direct imaging and LIM constraints simultaneously. This would be a very interesting exercise, but is well beyond the scope of the current analysis. 
Thus we will instead examine a few simple modifications to our SAM-based calculation and examine their implications.  We do not claim that either the results presented above or any in this section are definitive statements about the nature of the high-redshift universe, but they do offer a picture of the kind and scale of model-dependent effects that exist in this and future LIM analyses.

\subsection{Fixed $\rh$, varying $\alpha_{\rm{CO}}$}
As stated previously, our primary assumption when deriving Fig. \ref{fig:rhoH2} was that we know precisely the relationship between CO and molecular gas, so that an increase in CO intensity must be accompanied by an increase in $\rh$.  However, the same result could be accomplished by instead altering the relationship between CO and H$_2$.  The most obvious place to start would be the choice of $\alpha_{\rm{CO}}$.  We can hold the total $\rh$ constant and simply increase the CO luminosity per unit of molecular gas.  

In the SAM+sub-mm SAM, we know both the CO luminosity and H$_2$ mass of each simulated galaxy.  Thus, we can access the full distribution of $\alpha_{\rm{CO}}(M)$ for halos of different mass.  To compare to the values quoted above, we need the population-averaged value.  Averaging only over the star-forming population, this takes the form
\be
\overline{\alpha}_{\rm{CO}}=\frac{1}{\overline{n}}\int f_{\rm{duty}}(M) \alpha_{\rm{CO}}(M)\frac{dn}{dM} dM,
\ee
where $\overline{n}=\int f_{\rm{duty}}(M)(dn/dM)dM$ is the mean number density of star forming galaxies.

To fit the mmIME result by varying $\alpha_{\rm{CO}}$, we start with the end result of the previous section, a fitted value of $A^{\rm{SAM}}_{\rm{CO}}$ that brings the semianalytic prediction of $\Is$ into agreement with the mmIME measurement.  As we saw, this alone yields a $\rh$ estimate well in excess of that obtained through direct observations.  We then make a new assumption, that the W20 $\rh$ fit in fact provides the true value of the molecular gas density.  Finally, we linearly scale the SAM $\alpha_{\rm{CO}}$ parameter to a value which brings the red SAM+mmIME points from Fig. \ref{fig:rhoH2} into agreement with the W20 fit.  Symbolically, this is given by
\be
\alpha_{\rm{CO}}^{\rm{fit}}=\alpha_{\rm{CO}}^{\rm{SAM}}\frac{\rho_{\rm{H2}}^{\rm{W20}}}{A_{\rm{CO}}^{\rm{SAM}}\rho_{\rm{H2},0}}.
\ee

\begin{figure}
\centering
\includegraphics[width=\columnwidth]{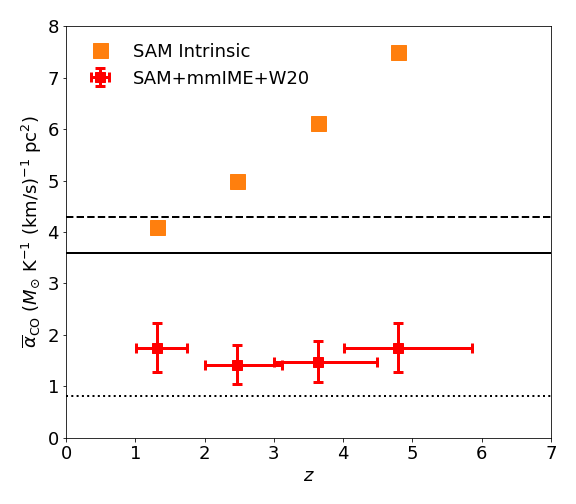}
\caption{Population averaged conversion constants $\alpha_{\rm{CO}}$ between CO(1-0) luminosity and molecular gas mass for the four mmIME redshift bins.  The black lines show the values assumed in K20 for all galaxies (solid, the default value used in their analysis), the Milky Way (dashed), and for ULIRGS (dotted).  Orange squares show the values from the SAM alone, red squares show the value needed to bring the SAM+sub-mm SAM into agreement with both the mmIME $\Is$ and W20 $\rh$ measurements.}
\label{fig:aCO}
\end{figure}

The results of this computation are plotted in Figure \ref{fig:aCO}, with a comparison both to the redshift-independent range of $\alpha_{\rm{CO}}$ assumed in K20 and the mean values obtained from the unmodified SAM+sub-mm SAM.  Note that, because the intrinsic SAM+sub-mm SAM predictions for $\rh$ are fairly close to the W20 fit, this is similar to the result we would obtain simply by scaling $\alpha_{\rm{CO}}^{\rm{SAM}}$ by $1/A_{\rm{CO}}^{\rm{SAM}}$ to directly cancel out the effect of increasing the CO luminosity.

We find that, under these assumptions, we would need an $\overline{\alpha}_{\rm{CO}}$ value of $\sim1.5-2\ M_{\odot}$ (K km s$^{-1}$ pc$^{2}$)$^{-1}$, roughly a factor of two lower than that assumed in K20.  The $\sim3.5\sigma$ offset between the red points and black line in Fig. \ref{fig:aCO} is again the same initial tension in $\Is$, but this time attributed entirely to a change in CO emission per unit of molecular gas.  

Note that throughout this section we have continued to hold all aspects the SAM+sub-mm SAM predictions to their intrinsic values with only two exceptions, the linear $A^{\rm{SAM}}_{\rm{CO}}$ scaling on $L_0(M)$ needed to fit the mmIME measurement, and a linear scaling on $\alpha_{\rm{CO}}$ to bring the resulting $\rh$ fit into agreement with W20.  The shape of $L_0(M)$, the scatter around that average, and the conversions between CO(1-0) and each of the higher transitions are all unchanged.

\subsection{Varying the sub-mm SAM subgrid ingredients}\label{sec:Plummer}

The sub-mm SAM and the underlying Santa Cruz SAM are both highly sophisticated frameworks, with many tunable parameters, necessary to model the complexity of the interstellar medium.  For the same reason that we do not attempt any fully automated parameter search, an exhaustive study of all of the possible SAM and sub-mm SAM variations that could affect $\Is$ is beyond the scope of this paper.  We will instead highlight a single change to show how sensitive the interpretation is to the choice of model.  Specifically, we will alter the density profile assumed for individual molecular clouds from the power-law shape we have assumed thus far to the Plummer density profile originally suggested by \cite{Popping2019}.  

\begin{figure}
\centering
\includegraphics[width=\columnwidth]{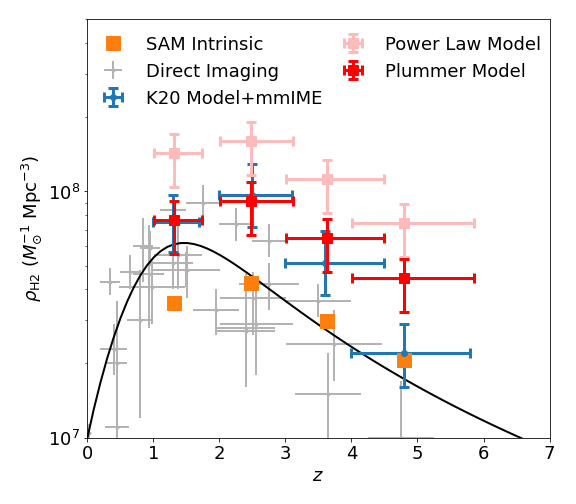}
\caption{Comparison of molecular gas mass density inferred from the mmIME measurement using a Plummer molecular cloud profile (dark red) to results from Fig. \ref{fig:rhoH2} (pink) which assumes a power law profile for the sub-mm SAM subgrid model.}
\label{fig:plummer}
\end{figure}

Using a new galaxy catalog produced under this changed assumption, we obtain a predicted $\Is$ of 198.7 $\mu$K$^2$ Hz sr, a factor of three or so higher than our original result.  The tension with the mmIME value of $1010^{+550}_{-390}$ $\mu$K$^{2}$ Hz sr is lessened, but still lingers at $\sim2\sigma$.  However, when we repeat the calculation of the cosmic molecular gas density from Fig. \ref{fig:rhoH2}, we see that the inferred $\rh$ values are now much closer to those obtained through the K20 scaling model, as shown in Figure \ref{fig:plummer}. We still see a modest increase for the highest-redshift CO(5-4) line, but as mentioned above, the highest redshift points are likely the least trustworthy under our limited ability to separate interlopers.  

As we have discussed in \S \ref{sec:SAM_intro}, both the power law and Plummer profiles produce results for the $L_\mathrm{CO}-\mathrm{SFR}$ relation that are in reasonable agreement with the observational constraints within the rather large scatter.  However, the $L_\mathrm{[CII]}-\mathrm{SFR}$ and $L_\mathrm{[CI]}-L_\mathrm{IR}$ relations predicted by the SAM $+$ sub-mm SAM assuming the Plummer model are significantly lower than the current observations. The purpose of showing the results with the Plummer radial profile assumption is not to suggest that it is the correct resolution to the tension we have pointed out. Instead, we show this variant to illustrate how sensitive the inferred $\rho_\mathrm{H_2}$ can be to details about the sub-grid properties of the ISM.

\subsection{Varying the K20 Scalings}

\begin{figure}
\centering
\includegraphics[width=\columnwidth]{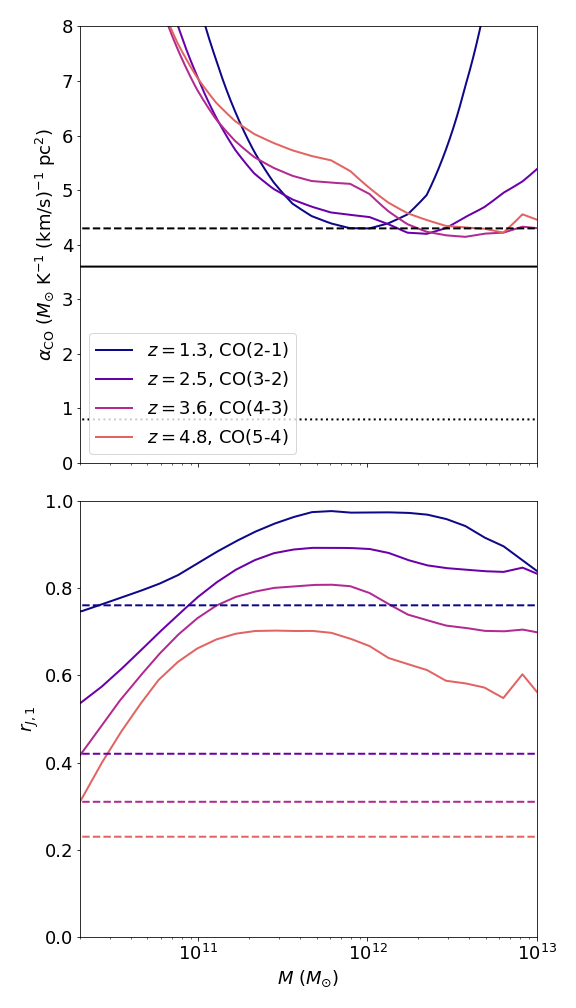}
\caption{Comparing the mass dependence of SAM scaling relations to those assumed in K20.  (Top) CO(1-0) luminosity--H$_2$ mass ratio $\alpha_{\rm{CO}}(M)$ at the central redshifts of our four CO transitions (colored lines) compared to the typical, Milky Way, and ULIRG values used in K20 (black solid, dashed, and dotted lines respectively).  All $\alpha_{\rm{CO}}$ averages are computed using only the star-forming galaxies from the SAM+sub-mm-SAM.  (Bottom) CO line luminosity ratios $r_{J,1}(M)$ for the same lines (solid lines) compared to the constant \citet{Daddi2015} values from K20 (dotted lines).  For both relations, note that the halo mass function tends to weight $10^{11}-10^{12}\ M_{\odot}$ halos the strongest when computing average quantities.}
\label{fig:ar}
\end{figure}

For the final step in our exploration, we will examine how the SAM+sub-mm SAM predictions for the individual scalings used in Section \ref{sec:scalings} differ from those assumed in K20, and estimate how these differences impact the estimate of $\rh$.  Though we did not explicitly use either the $\alpha_{\rm{CO}}$ or $r_{J,1}$ scalings to produce Fig. \ref{fig:rhoH2} from the SAMs, we can of course read these quantities directly from the simulated catalog.  The most obvious difference between the SAM-inferred relations and those in K20 is that both the CO-H$_2$ and line luminosity ratios are mass- and redshift-dependent.  Figure \ref{fig:ar} shows the two primary ratios used to convert the CO measurement to $\rh$, comparing the mass-dependent semi-analytic values to the constant empirical values assumed in K20.

We can immediately see clear and significant differences between the SAM predictions and the K20 assumptions.  For $\alpha_{\rm{CO}}$, the SAMs predict higher values at all masses and redshifts than the K20 assumption.  The $r_{J,1}$ values are generally higher as well, the redshift dependence is a bit different, and there is a significant halo mass dependence.  As the entire point of the SAM+sub-mm SAM is that all of these quantities are computed self-consistently based on a physical model, we cannot simply manually change these relations to see how they affect the measured $\Is$ --- any changes would have to by obtained by varying parameters like the cloud profile choice discussed in the previous section.  In order to get a sense for the role of different pieces of model for obtaining an estimate of $\rho_{\rm{H2}}$ from $\Is$, we need to artificially break the internal consistency of the semi-analytic predictions.  This will lose the key advantage of the SAM+sub-mm SAM procedure, but it will give us an idea of where the differences come from and how big they might be.

Combining all of the scalings from Section \ref{sec:scalings} allows us to explicitly write the dependence of $\rh$ on the fitted $A^{\rm{SAM}}_{\rm{CO}}$, yielding
\begin{multline}
\rh=A^{\rm{SAM}}_{\rm{CO}}\frac{1}{x'}J^3 \\ \times \int\frac{\alpha_{\rm{CO}}(M)}{r_{J,1}(M)}f_{\rm{duty}}(M)L_J(M)\frac{dn}{dM}dM,
\label{rhoscale}
\end{multline}
where $x'\equiv4.9\times10^{-5}\ L_{\odot}$ (K km s$^{-1}$ pc$^2$)$^{-1}$ is the conversion factor between $L$ and $L'$ from Eq. (\ref{Lprime}).  The factor of $J^3$ comes from the frequency dependence of the same conversion.  Since this final computation is even more approximate than the previous ones, we will neglect scatter in these quantities for now.  Eq. (\ref{rhoscale}) without scatter gives a $\rh$ result within 5\% of the original calculation using the full SAM+sub-mm SAM results with scatter.
We can see that the $\rh$ inferred from a given line scales as $\alpha_{\rm{CO}}/r_{J,1}$, up to an integral over mass.  We can modify this ratio to change $\rh$ without changing the assumed mass-luminosity relation $L_J(M)$.

\begin{figure}
\includegraphics[width=\columnwidth]{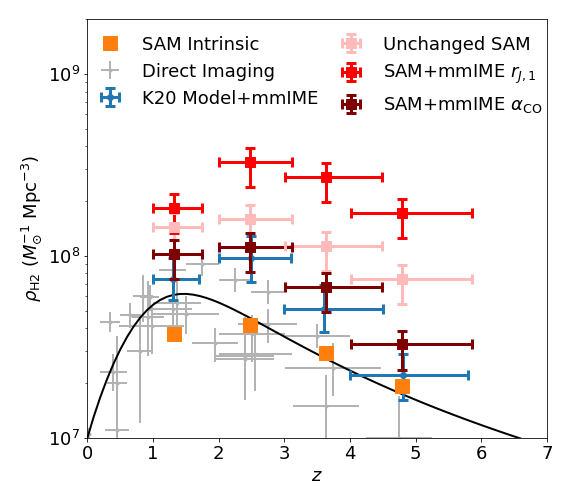}
\caption{Estimates of $\rh$ obtained by combining the semi-analytic and empirical scaling relationships in Eq. (\ref{rhoscale}), compared to the same K20 mmIME and direct-imaging constraints plotted in Figs. \ref{fig:rhoH2} and \ref{fig:plummer}.  Light red points show our original constraints which use the unmodified SAM+sub-mm SAM to go from $A^{\rm{SAM}}_{\rm{CO}}$ to $\rh$.  Red and dark red points respectively substitute in the constant line ratios $r_{J,1}$ and CO-H$_2$ ratios $\alpha_{\rm{CO}}$ values from K20, while retaining the other model elements from the SAM+sub-mm SAM predictions.}
\label{fig:arscale}
\end{figure}

To get a rough idea of how much this choice matters, in Figure \ref{fig:arscale} we replace alternately the $\alpha_{\rm{CO}}(M)$ and $r_{J,1}(M)$ relations in Eq. (\ref{rhoscale}) with the constant values used in the analysis of K20. If we use the commonly-assumed $\alpha_{\rm{CO}}=3.6\ M_{\odot}$ (K km s$^{-1}$ pc$^{2}$)$^{-1}$, our $\rh$ values fall by $1-1.5\sigma$, and end up much closer to the K20 values.  This effectively restates the point from Fig. \ref{fig:aCO} that a smaller $\alpha_{\rm{CO}}$ will alleviate the tension in our $\rh$ estimates.  On the other hand, if we consider ourselves free to modify the model in this fashion, keeping the built-in mass-dependent $\alpha_{\rm{CO}}$ and instead using the \citet{Daddi2015} constant $r_{J,1}$ ratios increases the  inferred molecular gas abundance even further beyond the estimates from direct imaging.  All in all, the range between the different estimates in Fig. \ref{fig:arscale} can be seen as an extremely approximate estimate of the ``error bar" on the modeling uncertainty in this problem, which is substantially larger than the statistical uncertainties.

We should note that this last type of uncertainty is far from unique to LIM measurements.  Direct-imaging measurements of CO transitions must also make assumptions about the values of $r_{J,1}$ and $\alpha_{\rm{CO}}$, and these choices are a source of significant uncertainty \citep[see][for an example]{Boogaard2021}.  One could thus envision a similar amount of model-based scatter on the gray literature points in Fig. \ref{fig:arscale} as on the red LIM points.  As discussed above, however, the need to choose mass dependent values for these quantities presents and will continue to present a unique challenge for integrated LIM measurements.

\section{Discussion}
\label{sec:discussion}

The core result presented here is both simple and fairly robust, at least at the level of detail we consider here.  The mmIME survey produced a measurement of the total shot power from high-redshift CO lines that is roughly $3.5\sigma$ in tension with predictions from the Santa Cruz SAM+sub-mm SAM framework, taken as face value.  The tension is present in both versions of the SAM+sub-mm SAM discussed above, which adopt different molecular cloud density profiles.  There are of course many other modeling choices baked in to the semi-analytic prediction that we have assumed here to be essentially fixed.  Most of these parameters, however, are set by comparison to a wide variety of other galaxy survey measurements, and changing any of them to fit the mmIME data could easily create discrepancies elsewhere.  A full statistical treatment would require fitting the mmIME $\Is$ value and all of the other galaxy measurements simultaneously while varying all of the underlying SAM parameters.  Such an effort would be highly enlightening, but is beyond the scope of this work, so for this initial exploration we confine ourselves to reproduction and modest extension of the original K20 mmIME analysis using our semi-analytic framework.

When we repeat an analysis similar to that of K20, but replace their empirical scalings with predictions from the SAM for the relationship between halo mass and CO luminosity, and assume that any difference between the mmIME measurement and our model prediction of $\Is$ can be explained by a linear change in H$_2$ content across all halo masses, we infer the presence of a large amount of excess molecular gas at high redshift compared to direct measurements from individual CO line detections or dust continuum measurements.  If true (which is far from certain given the limitations of these results), the distribution of H$_2$ within galaxies would almost certainly not look like a simple linear increase in every galaxy above the original SAM+sub-mm SAM predictions.  In order to have been missed in traditional surveys, the excess molecular gas would have to be predominantly located in individually-faint lower mass galaxies, otherwise surveys like COLDz and ASPECS would see it (though there may be sample variance limitations as well, see \citealt{Keenan2020}).  If tensions like this continue to exist and are confirmed in the future, a simple extension to this analysis would be to apply something like the empirical models presented in \citet{Padmanabhan2018}, where we are free to add more CO emission to galaxies of different masses rather than uniformly increasing it across the board as was done here.

The mmIME measurement is not the only source of tension between intensity mapping measurements and semi-analytic forecasts.  \citet{Yang2019} published an estimate of the 158 $\mu$m \ion{C}{2} integrated line intensity by cross-correlating data from the Planck satellite with spectroscopic galaxy measurements.  As shown in Figure 8 of Y20, the result is also somewhat brighter than the prediction from the SAM+sub-mm SAM.  While this tension is also quite modest, it is further evidence that some of the physical processes in the current Santa Cruz SAM + sub-mm SAM may need to be modified. 

As emphasized previously, there are many uncertainties in the physical processes and parameter values for both the SAM and sub-mm SAM. There are also other signs of observational tensions with the Santa Cruz SAMs and other models --- for example, \citet{2019ApJ...882..137P} showed that both the Santa Cruz SAM and the IllustrisTNG hydrodynamic simulations show a deficit of large H$_2$ reservoirs at intermediate redshift $z\sim 1-2$. This could reflect inadequacies in the modeling of processes such as star formation and stellar feedback in the galaxy formation models, or the fairly simple assumptions about conversion from the observed CO lines to $H_2$ mass adopted in that work. As already discussed, addressing a disagreement with a particular set of observations by varying the physics in the models is tricky, because the SAM has been developed over many years to produce agreement with a broad set of observations over a wide range in redshift. Ultimately, a more automated calibration procedure needs to be developed that can account for many observations simultaneously.  Similarly, there are many assumptions that go into the sub-grid models that need to be adopted to describe the ISM properties and carry out the line spectral synthesis. The properties of the ISM such as the cloud mass function, and cloud radial profiles, may vary in a complex way with environment or other galaxy properties. Again, as already noted, changing one of the subgrid ingredients to fit a particular line may result in worse agreement for a different line arising from a different part of the ISM, as we found when changing the cloud radial profile from a power law to a Plummer profile. 

Key to resolving all of these challenges will be of course acquiring more data.  Fortunately, there will be a dramatic increase in the volume and quality of LIM data in the next few years.  The Carbon Monoxide Mapping Array Project \citep[COMAP][]{Li2016} targets the CO(1-0) emission line at $z=2.4-3.4$, right in the middle of the mmIME redshift range and near the excess CII emission seen in the Planck data, and would provide a highly complementary molecular gas measurement over a much larger intensity mapping volume.  The Experiment for Cryogenic Large-Aperture Intensity Mapping \citep[EXCLAIM][]{Cataldo2021} will probe the CII line in a similar redshift range, and will offer an improved measurement of that line to improve upon the Planck measurement.  Several other reionization-focused projects have a mmIME-like ladder of CO lines as a ``foreground" to their higher-redshift science \citep{Lagache2018,Stacey2018,Sun2020}.

These new experiments will offer opportunities for more varied analysis of the intensity mapping data beyond the power spectra used in mmIME.  Some surveys, including COMAP and EXCLAIM, are designed with the possibility in mind of cross-correlating their intensity maps with spectroscopic galaxy surveys.  This has the primary utility of improving confidence in the detection of a cosmological signal, but cross-correlations can also provide windows into unique aspects of galaxy evolution beyond those accessible to a single line \citep{Serra2016,Breysse2017,Breysse2019,Chung2019}.  Even without cross-correlations, other quantities like the anisotropic power spectrum \citep{Lidz2016,Bernal2019,Chung2019a,Cheng2016} and the one-point intensity distribution \citep{Breysse2017a,Ihle2019,Breysse2019a} permit more detailed probes of the distribution of line intensity than the single shot power measurement available in mmIME.

\section{Conclusion}
\label{sec:conclusion}

We have demonstrated here that a modest tension exists between new line intensity mapping observations of high-redshift galaxies from the mmIME survey and state-of-the-art semi-analytic models of the line emission from the dense ISM in galaxies.  At $\sim3.5\sigma$, this tension is relatively minor and may weaken with time and new data.  However, these are some of the first intensity mapping results at these redshifts, and the SAM+sub-mm SAM routines have been calibrated to date primarily on direct galaxy survey measurements, so even this modest tension is interesting.  We therefore went on to explore the ways in which this tension could be caused by physical processes that impact the LIM measurements.

We carry out a modified version of the K20 analysis of the mmIME-based measurement of the cosmic molecular gas abundance, using our internally-consistent, physically motivated SAM+sub-mm SAM scaling relations at all steps.  In doing so, we found implied $\rho_{\rm{H2}}$ values roughly twice as high as those obtained in K20, corresponding to a $\sim2\sigma$ increase.  As the K20 $\rho_{\rm{H2}}$ fit is already a bit higher than measurements obtained from direct imaging galaxy surveys, our result is quite a bit higher than any previous measurement.  In other words, if we assume, as in K20, that any tension between mmIME and the model predictions is due to a change in the molecular gas abundance, the modest tension in the intensity map becomes a similar tension in $\rho_{\rm{H2}}$. If this interpretation is valid, it implies that there is a substantial population of CO-emitting galaxies that are statistically detectable to LIM but have not been seen in galaxy surveys.

However, we go on to show that the tension could similarly be resolved by a number of other changes in the modeling assumptions.  Most simply, we show that we could equivalently account for the tension by reducing the value of $\alpha_{\rm{CO}}$ in high-redshift galaxies, without changing the molecular gas abundance at all.  We also show that small changes to the semi-analytic model sub-grid ingredients can solve the tension in $\rho_{\rm{H2}}$ even if the overall tension with mmIME remains.  Finally, we relaxed our reliance on the SAM+sub-mm SAM slightly and showed that substituting some of the not-unreasonable scaling relations originally used in K20 can shift our $\rho_{\rm{H2}}$ result over a range quite a bit larger than the statistical error bars on the measurement, with some choices weakening the tension and others leading to even higher $\rho_{\rm{H2}}$ values.

It is thus clear that, while current data and model predictions have perhaps begun to differ in interesting ways, neither has progressed to the point of being able to confidently understand the source of those differences.  This state of affairs is rapidly changing however, particularly on the experimental side.  Intensity mapping data in particular will dramatically expand in the next few years, as many dedicated LIM instruments begin to produce results.  Perhaps the most important conclusion of this work is that substantial effort will be required on the modeling and simulation fronts if we are to fully understand the new regimes probed by these measurements.

\section{Acknowledgements}
We thank Garrett Keating, Dongwoo Chung, Hamsa Padmanabhan, and H\"avard Ihle for usefull discussions and comments on this manuscript.  This work made use of the Flatiron Institute Computing Cluster.  PCB was supported by the James Arthur Postdoctoral Fellowship. ARP was supported by NASA under award numbers 80NSSC18K1014 and NNH17ZDA001N.  RSS gratefully acknowledges support from the Simons Foundation.

\bibliography{references}{}
\bibliographystyle{aasjournal}

\end{document}